\documentclass[12pt,a4paper]{article}
\usepackage[latin1]{inputenc}
\usepackage{amsmath}
\usepackage{amsfonts}
\usepackage{amssymb}
\usepackage{lscape}
\usepackage{hyperref}
\hypersetup{
    colorlinks=true,
    pdfborder={1 1 1},
} 
 \usepackage{graphicx}

\newcommand{\half}{  {\scriptstyle \frac{1}{2} } }

\newcommand{\sF}{{\cal F}}
\newcommand{\sG}{{\cal G}}
\newcommand{\sB}{{\cal B}}
\newcommand{\sS}{{\cal S}}
\newcommand{\h}{h}

\newcommand{\wl}{ \underline{w} }
\newcommand{\wh}{ \bar{w} }

\newcommand{\Bl}{\underline{B}}
\newcommand{\Bh}{\bar{B}}
\newtheorem{lemma}{Lemma}

\author{Moritz D\"umbgen
and L.~C.~G. Rogers\\
University of Cambridge}
\title{Investing and Stopping}
\begin{document}
\maketitle

\begin{abstract}
In this paper we solve the hedge fund manager's optimization problem
in a model that allows for investors to enter and leave the fund over time
depending on its performance. The manager's payoff at the end of the 
year will then depend not just on the terminal value of the fund level, 
but also on the lowest and the highest value reached over that time. 
We establish equivalence to an optimal stopping problem for Brownian
motion; by approximating this problem with the corresponding optimal
stopping problem for a random walk we are led to a simple and efficient
numerical scheme to find the solution, which we then illustrate 
with some examples.
\end{abstract}

\section{Introduction.}\label{S0}
The fee structure of a hedge fund typically consists of two 
components, a fixed management fee\footnote{This will usually
be a relatively low percentage, 2\% being common.},
 charged on all assets under management, and a performance 
 fee\footnote{This is usually charged at quite a high rate, 
 20\% being common.}, charged on any gain achieved on the funds 
 invested. The exact contractual agreement has to specify between
 what dates the gain must have been recorded, what happens to the 
 management fee for funds deposited for part only of a period of 
 reckoning, and many other details, such as any restrictions on 
 investors' freedom to withdraw funds with or without notice
 periods. We shall simplify the problem here, by assuming that
 the performance fees are charged at the end of each year on all
 funds held at the end of the year\footnote{It may be that some
 funds are withdrawn before the end of the year, and could in principle
 be liable to pay performance fees, but we shall ignore this, on 
 the grounds that investors would be unlikely to withdraw funds
 while they were ahead.}, and the gain is calculated as the increase
 in value of the funds from the time they were deposited in the
 the fund, or from the beginning of the year, whichever is later.
 Thus the baseline for calculating the performance fee resets at the
 beginning of each year. We shall suppose that the management fee is 
 charged only on the funds still under management at the end of the
 year; this is a simplification, but as the management fee is typically
of  smaller magnitude, it is relatively innocent.

In principle, the total fees charged at the end of the year by 
the hedge fund to its clients would depend on the entire history of
investments and withdrawals through the year, as well as on the 
actual performance path of the hedge fund. We shall propose a 
simplified mechanism for this, which involves some story about
how the quantity of assets under management varies as the level of 
the hedge fund fluctuates, and is explained in detail in Section
\ref{S2}. This story captures the key features that the AUM 
rise as the level of the fund rises, and fall as the level falls;
that newly-invested funds enter at the current level; and
that funds withdrawn will have entered the fund at a level above 
the current level.  The story we tell is not perfect, but has the
crucial simplifying property that {\em
the fees paid will depend
on the level of the fund at the start of the year, at the 
end of the year, and on the highest and lowest levels attained.}
This saves us from the need to carry along as a state variable the
entire profile of the levels at which the current AUM entered the
fund, which then would be impossibly clumsy to work with (compare
with the study of Dybvig \& Koo \cite{dk96} on wash sales). 

In a seminal contribution to this subject, Goetzmann, Ingersoll and Ross \cite{gir03} provided closed-form solutions
to a model which differs from ours fundamentally in that the performance fee is considered to be paid out continuously over time (with a high-water mark provision). This has the undesirable side-effect that at the end of the year, the manager's reward is a function only of the high-water mark. Guasoni and Ob\l oj \cite{go10} followed a similar approach with a continuously paid performance fee, but modelled the manager as a utility optimiser himself, also resulting in (asymptotic) closed-form solutions.

Accepting this simplified model, we find ourselves with an optimal
control problem for the hedge fund manager, in which the objective
is a function of the initial, final, highest and lowest values 
taken by the controlled process in the year.  We shall suppose that
the riskless rate is zero, and that the hedge fund manager is in
fact just investing in assets which fluctuate but have no drift.
This ignores a fund manager's presumed ability to pick winners,
time the market, anticipate roll dates, or any  other 
marketing boast; this may be unjust at the level of a single 
talented manager, but not too far from the situation for the 
industry as a whole.  
An alternative justification is that while the assets invested in 
might have a positive drift, the manager will take expectations
under an equivalent measure which removes the drift, as a
risk control measure.

The level of the fund will therefore 
evolve in our model  as a martingale, which for simplicity we suppose is 
continuous; the manager can adjust the volatility of the level process
by choosing a smaller or larger position in the risky assets, but
he cannot affect the drift. Nevertheless, he has an incentive to 
embrace some risk, as he has a call option interest in the level of the
fund, as well as the performance fee incentive.
  Since any continuous martingale is a time change of Brownian 
motion, we shall begin our analysis by converting the manager's problem
into an optimal stopping problem for a Brownian motion. This is 
not quite as obvious a step as might at first sight appear, as we 
explain in Section \ref{FE}.
 The next step is to convert the optimal stopping
problem for Brownian motion into an optimal stopping problem for 
a symmetric simple random walk (SSRW), whose value will be close to the 
value of the original problem; the difference is analyzed and 
estimated in Section \ref{SE}.  While it would be possible to write
 down some 
formulation of the solution to the original continuous problem, it 
would not be particularly digestible, and there would then be the 
issue of existence and uniqueness of solutions. Since we do not 
expect ever to be able to exhibit any closed-form solution, we 
are forced to numerical methods to gain understanding; and these
are naturally discrete in nature. Our estimates allow us to be quite
precise about the error committed by the approximation.
 Finally, this problem can be solved quite 
efficiently numerically, as we demonstrate in Section \ref{SP}.

We then return in Section \ref{S3} to the hedge fund manager's problem,
where we state our modelling assumptions on how cash flows into and 
out of the fund as the level of the fund varies, converting the manager's
objective into one of the type studied in Section \ref{S2}. We then present
numerical solutions of this problem. Section \ref{S4} concludes.

\section{From investing to stopping.}\label{S2}
In this Section, we firstly show that the investment problem 
can be recast as a stopping problem for Brownian motion; then we
show that this stopping problem can be approximated by the 
corresponding stopping problem for SSRW;
and finally we explain the algorithm for solving this SSRW stopping
problem.

\subsection{The investing/stopping equivalence.}\label{FE}
We suppose that the level of the fund is $w_0$ at time 
0, and evolves as 
\begin{equation}
	dw_t = \theta_t dW_t \label{dw}
\end{equation}
for some previsible process $\theta$ for which the stochastic
integral is defined, where $W$ is a standard Brownian motion.
We define
\begin{equation}
	\wl_t \equiv \inf\{ w_s: s \leqslant  t\}, \qquad \wh_t \equiv \sup\{w_s: s \leqslant  t \},
	\label{wlh}
\end{equation}
and we suppose that the objective of the manager of the fund is 
\begin{equation}
	\sup_\theta\, E  F(\wl_1, w_1, \wh_1)\label{obj1}
\end{equation}
for an $F\!:\!(-\infty, w_0]\times \mathbb R \times [w_0, \infty)\rightarrow \mathbb R$ 
for which the above expectation is always well-defined (continuous and bounded from above or below, say). The well-known 
Dubins-Schwarz result says (informally) that any continuous local 
martingale is a time-change of a Brownian motion.  More precisely, 
if we extend the definition \eqref{dw} of $w$ beyond time 1 by setting 
$\theta_t = 1$ for all $t \geq 1$, and set   
\begin{equation}
	A_t \equiv \int_0^t \theta_s^2\;ds, \qquad \tau_t \equiv \inf\{ s: A_s >t\}, \label{Atau}
\end{equation}
then $B_t \equiv w(\tau_t)$ defines a Brownian motion relative
to the filtration $\sG_t \equiv \sF_{\tau_t}$, and each $A_t$ is 
a $\sG$-stopping time.  It follows directly that
\begin{equation}
	w_t = B(A_t), \qquad 
	\wl_t = \inf_{0 \leqslant  s\leqslant  A_t} B_s \equiv \Bl(A_t), \qquad  
	\wh_t = \sup_{0 \leqslant  s \leqslant A_t} B_s \equiv \Bh(A_t),\label{DS1}
\end{equation}
As a consequence, {\em were it not for the fact that $A_1$ is not in 
general a stopping time for $B$,} the following result would be 
trivial.

\begin{lemma}\label{L1}
	For a continuous $F$, bounded from above or below, the equality
	\begin{equation}
		\sup_\theta\, E  F(\wl_1, w_1, \wh_1)  = \sup_{T \in {\cal T} }\, E F(\Bl_T, B_T,\Bh_T) \label{eq8}
\end{equation}
is valid, where ${\cal T}$ denotes the set of stopping times of 
the Brownian motion $B$.
\end{lemma}

\bigskip\noindent
{\sc Remarks.} Suppose that $M$ is a continuous martingale which
runs like a Brownian motion until some independent exponential
random time
$T$, then stands still for one unit of time, and then resumes 
Brownian motion. The quadratic variation process $[M]$ grows
at rate 1 except in the interval $[T,T+1]$, where it remains
constant.  It is quite easy to show that
$[M]_T$ is a $\sG$-stopping time, but it is impossible to discover
what $[M]_T = T$ was just by looking at the time-changed 
Brownian path $B_t = M(\tau_t)$. 
 Thus we expect the 
left-hand side of \eqref{eq8} to be at least as big as the 
right-hand side, but it is not initially obvious that the two sides
are the same.

\bigbreak\noindent
{\sc Proof of Lemma \ref{L1}.}  
See Appendix.

\subsection{Approximation by random walk.}\label{SE}

Thanks to Lemma \ref{L1}, we are now left to solve an optimal stopping 
problem for a Brownian motion whose stopping reward is a function of its 
current value, minimum and maximum, 
\begin{align}\label{stopbm}
	V			=\sup_{T \in {\cal T} }\, E F(\Bl_T, B_T,\Bh_T).
\end{align}
Although suppressed in the notation, we think of $V$ as a function of the starting 
values $X_0 \equiv (\Bl_0, B_0, \Bh_0)$. It should come as no surprise 
that we can approximate 
$V$ uniformly by a stopping problem for a SSRW $w^\h$ 
on the grid $B_0 + \h \mathbb Z$,
\begin{align}\label{stoprw}
	V^\h	=\sup_{T \in {\hat{ \cal T}} }\, E F(\underline w^\h_T, 
	w^\h_T, \bar w^\h_T),
\end{align}
where $\hat{ \cal T}$ represents all (discrete) $w^\h$-stopping times.

\begin{lemma}\label{L2}
Let $F$ be uniformly continuous: there is some continuous function 
$\psi$ tending to zero at zero  such that for all $x, x'$
\begin{equation}
|F(x) - F(x')| \leqslant\psi( \Vert x-x' \Vert).
\nonumber
\end{equation}
If the optimization problem is well posed, then
\begin{equation}
| V^h(X_0) - V(X_0)| \leqslant\psi(\h  \sqrt{3} ).
\end{equation}
\end{lemma}
 
 \noindent{\sc Proof of Lemma \ref{L2}.}  
See Appendix.

\subsection{Solving the random walk stopping problem.}\label{SP}
Given that we have now replaced the original investment problem with
an optimal stopping problem for a SSRW, we are in a position to solve
it by numerical means\footnote{It is inconceivable that we may be able to 
find closed-form solutions, except in some
very contrived examples.}.
If we are to follow this route, then we will of course only be able to 
deal with examples which are finite, and for this reason we are
justified in assuming that the random walk will be stopped once it leaves
some interval $(w_*,w^*)$ containing $w_0$.
\begin{figure}[h!]
	\centering
 	\includegraphics[scale=0.8,angle=0]{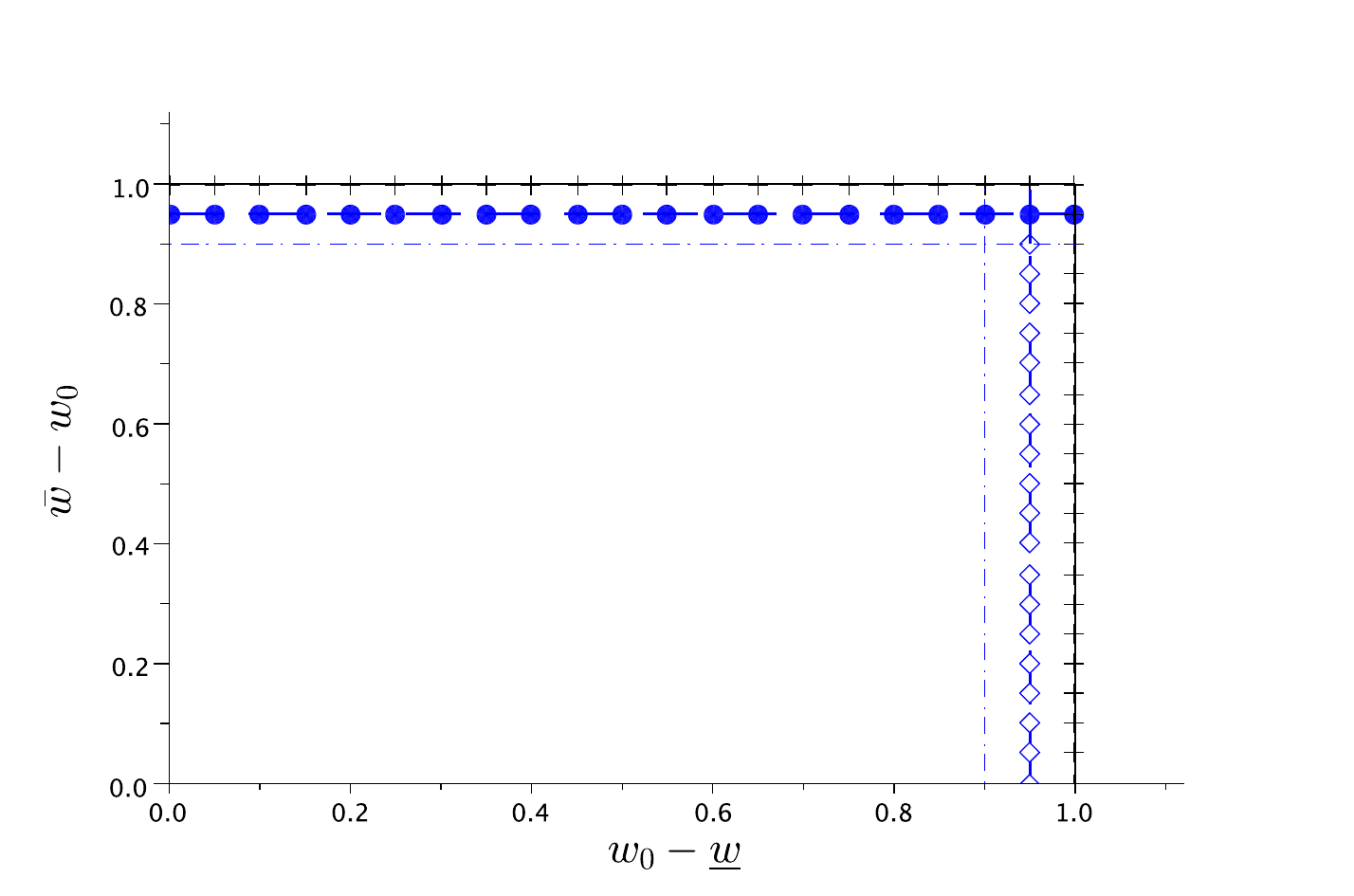}
	
	Figure A: Filling in the value function. Here, $w^*-w_0=w_0-w_*=1$.
\end{figure}
Now suppose that $w_* = w_0 - m\h$, $ w^* = w_0+ n\h$ for some positive
integers $m,n$, and introduce the notation
\begin{equation}
F_{jk}(i) = F( w_0 - j\h, w_0 +i\h, w_0 + k\h),\quad
V_{jk}(i) = V^\h( w_0 - j\h, w_0 +i\h, w_0 + k\h)
\nonumber
\end{equation}
for $-m \leqslant -j \leqslant i \leqslant k \leqslant n$. We have
that $V_{jk} \geqslant F_{jk}$ always, and that if $j = m$ or 
$k=n$ equality holds, since the random walk must have stopped by the 
time it reaches those points. We can now solve recursively for the 
value function $V$ rather as we would solve a dynamic programming
problem. We shall have that for $-j < i < k$
\begin{equation}
V_{jk}(i) = \max\{ F_{jk}(i), \half( V_{jk}(i+1) + V_{jk}(i-1))\, \}
\label{BE1}
\end{equation}
and at the ends of the interval we have
\begin{eqnarray}
V_{jk}(-j) &=& \max\{ F_{jk}(-j), \half( V_{jk}(-j+1) + V_{j+1,k}(-j-1))\, \}
\label{BElo}
\\
V_{jk}(k) &=& \max\{ F_{jk}(k), \half( V_{j,k+1}(k+1) + V_{jk}(k-1))\, \}.
\label{BEhi}
\end{eqnarray}
The situation is illustrated in Figure A, where we plot the grid
of $(\wl,\wh)$ pairs, and may imagine that we are looking down on a cube, 
each point of the form $(w_0 - j\h, w_0 + k\h)$ being the projection
down into the plane of points of the form $(w_0-j\h, w_0 + i\h, w_0+k\h)$,
$\; -j \leqslant i \leqslant k$.
At every point of the upper right boundary of the rectangle, where
either $j = m$ or $k = n$, the value function is equal to $F$ and is therefore
known. Now we work out the values $V_{m-1, n-1}(i)$, by  solving the
optimal stopping problem \eqref{BE1} with the boundary conditions \eqref{BElo}
and \eqref{BEhi}. The two boundary conditions require knowledge of 
$V$ at $(m-1,n)$ and at $(m,n-1)$; but these values are then known, since
we know $V=F$ on the solid upper right boundary of the rectangle.
Now we calculate the value of $V_{m-2,n-1}$; this time, we need to know 
$V$ at $(m-2,n)$ - where it agrees with $F$ - and at $(m-1,n-1)$  - which
we calculated at the first step. Continuing in this fashion, we are able to 
calculate the values of $V$ at all points of the form $(\ell, n-1)$, 
represented by big dots in Figure A.  In like fashion, we can then 
work out the values of $V$ at all points $(m-1, \ell)$
marked with diamonds, which gives us the
values of $V$ not only at the upper right boundary, but at the the next layer
in, depicted by the dots and diamonds in the diagram.
 But now we have reduced the
size of the rectangle by one in each direction, so we can repeat the 
method just explained to find all the values of $V$ on the dot-dash lines.
Proceeding similarly gives us the solution $V$.

\bigskip\noindent
{\sc Remarks.}   At each node $(j,k)$ of the rectangle in Figure A
we have to solve an optimal stopping problem for random walk in
$\{ i : -j-1 \leqslant i \leqslant k+1 \}$, where the random walk
is absorbed at the endpoints $-j-1$ and $k+1$, with values 
$V_{j+1,k}(-j-1)$ and $V_{j,k+1}(k+1)$ respectively, and with 
stopping values $F_{jk}(i)$ at interior points. The value has a
geometric interpretation as the least concave majorant of the 
function defined by the stopping values, and can be calculated rapidly
and accurately by  policy improvement.

It may happen that for a given $(j,k)$ the optimal stopping solution 
is not to stop in the interior, but otherwise there will be a 
smallest value $w_0+q \h = \eta_l(j,k)$ and a largest value 
$w_0+\ell \h = \eta_u(j,k)$ at which $V_{jk}(i) = F_{jk}(i)$.  By convention,
we may define $\eta_l(j,k) =w_0+ (k+1)\h$ and $\eta_u(j,k) = w_0-(j+1)\h$
 if the optimal stopping solution does not admit stopping in 
  the interior of the interval.  At times $\tau$  when the random walk
  reaches a new maximum $w_\tau = \wh_\tau = w_0 + k\h$, it will
  thereafter continue until either it hits $w_\tau+\h$, or it hits
  $\eta_u(j,k)$, where $\wl_\tau = w_0 - j\h$.  If it hits the lower 
  barrier $\eta_u(j,k)$ before it hits $w_0+(k+1)\h$, then it will stop
  there for good,
  {\em unless} $\eta_u(j,k)=w_0 - (j+1)\h$, in which case a new
  minimum has been achieved, and the random walk can continue to move.

\section{The Hedge Fund Manager's Investing Problem.}\label{S3}
We return to the problem introduced in Section \ref{S2} of the hedge
fund manager, who can control the level $w_t$ of the fund through 
the position $\theta_t$ in  the risky asset.  As the level of the 
fund goes up and down, the assets under management vary. We propose
a very simple story for this which allows us to represent the 
manager's problem in the form \eqref{obj1}, which can then be solved
by the techniques just presented.

\medskip
The basic idea is that there is some $C^1$ non-negative function
$\varphi$ such that at any time $\tau$ when the level process $w$ is 
at its running maximum, $w_\tau = \wh_\tau$, the profile of the basis levels
of the assets in the fund should be given by 
\begin{equation}
    \varphi(x) dx, \qquad  0 \leqslant x \leqslant \wh.
    \label{profile}
\end{equation}
So in particular, the total assets under management at $\tau$ would be
$\Phi(\wh) \equiv \int_0^{\wh}  \varphi(x) \; dx$. If  we demanded that
$\varphi$ was increasing, this would represent a situation where the 
more successful the fund, the more people would bring their money to it.

What happens as the level of the fund falls back from its running maximum?
Investors will take their money out of the fund; as the level rises again,
investors will put money in.  Now as the level rises again and new
money comes into the fund, the basis at which that new money was invested
has to be {\em the current level.} In order to retain tractability, we shall
insist that when money is withdrawn from the fund as the level falls,  
{\em it is  removed only at the current level.}  This is a restrictive 
assumption,  but we make it nevertheless. So as the
level falls, a fraction $(1-p)$ of the assets invested at the current
level are removed from the fund, so that in general the profile of 
basis values in the fund is
\begin{equation}
\varphi(x)I_{\{x \leqslant w_t\}} + p \varphi(x)I_{\{w_t
\leqslant
x \leqslant \wh_t\}},
\label{prof2}
\end{equation}
which is consistent with \eqref{profile} when $w_t = \wh_t$.  The assumption
we make means that if the level of the fund falls a long way, there will 
be still quite a lot of assets which came in at a higher level, and have
not been taken out yet. This could be understood in terms of the reluctance
of investors to realize a loss; investors would be willing to come out at
zero gain, and they do in our story, but they would never come out if
they would thereby realize a loss. 

This is not the whole story, because the performance-related part of the 
manager's fees will be measured relative to the level $w_0$ of the fund
at the start of the year.  We shall therefore suppose that 
initially the profile of basis levels is a point mass at $w_0$ of
magnitude $\Phi(w_0)$, and that if the level falls to $\wl$ then 
the funds $(1-p)(\Phi(w_0)-\Phi(\wl))$ which would be removed if 
the profile $\varphi$ extended through $(0,w_0)$ will be removed from the 
atom at zero. Thus when the minimum value of the level is $\wl$, the
size of the atom at $w_0$ will be
\begin{equation}
   \Phi(w_0) - (1-p)(\Phi(w_0)-\Phi(\wl))
   = p \Phi(w_0) + (1-p) \Phi(\wl).
   \label{atom}
\end{equation}
Thus overall the profile of the basis levels will be
\begin{equation}
\bigl[\;
\varphi(x)I_{\{\wl_t \leqslant x \leqslant w_t\}} + p \varphi(x)I_{\{w_t
\leqslant
x \leqslant \wh_t\}}
\; \bigr] dx + \bigl\lbrace
p \Phi(w_0) + (1-p) \Phi(\wl)
\bigr\rbrace   \delta_{w_0}.
\label{final_profile}
\end{equation}
Integrating this gives the total assets under management as
\begin{eqnarray}
\textrm{AUM} &=& \Phi(w) - \Phi(\wl) + p( \Phi(\wh) - \Phi(w)) 
+ p \Phi(w_0) + (1-p) \Phi(\wl)
\nonumber
\\
&=& (1-p)\Phi(w) + p (\Phi(\wh) -\Phi(\wl)) + p \Phi(w_0).
\label{AUM}
\end{eqnarray}
We shall suppose that there is some constant $\beta \in (0,1)$ such that 
the manager receives
\begin{equation}
\textrm{MF} 
=\beta\times  \textrm{AUM}
\label{MF}
\end{equation}
 as the management fee.
The profile \eqref{final_profile} allows us to calculate the performance
component of the manager's reward, which will be 
\begin{equation}
\textrm{PF} = 
\alpha \bigl[ \;
\int_{\wl}^w (w- x) \varphi(x) \; dx + (w- w_0)^+
 \bigl\lbrace
p \Phi(w_0) + (1-p) \Phi(\wl)
\bigr\rbrace   
\; \bigr].
\label{PF}
\end{equation}

\subsection{Numerical examples.}\label{numerics}
We suppose that the manager is risk averse, so he tries to 
maximize
\begin{equation}
E U(\textrm{MF + PF}).
\end{equation}
In this first example, we take $U=\log(x)$, $\varphi = \sqrt{x\wedge K}$, $\alpha=20\%$, $\beta=2\%$, $p=0.3$, $w_0=1$, $\Phi(w_0)=1$ and $K=3$. Figures \ref{Fig1} and \ref{Fig2} illustrate the resulting payout function $F=U(\text{MF}+\text{PF})$ as a function of the three variables $(\wl, w, \wh)$. Since it is decreasing in $\wl$, increasing in $\wh$, as well as S-shaped in $w$ (first convex, then concave), we expect non-trivial results from the stopping problem. 

It is worth understanding why this should be. For fixed $(\wl, \wh)$, the optimal stopping problem is on the grid $\{\wl\!-\!h, \wl, ..., \wh, \wh\!+\!h\}$, where if we stop at $x\in[\wl, \wh]$, we get reward $F(\wl, x, \wh)$, but if we stop at one of the endpoints, we get (at the upper endpoint for example) $F(\wl, \wh+h, \wh+h)$. This value can be (and in places is) significantly bigger than $F(\wl, \wh, \wh)$, so we see a picture like Figure B. But if the values at the endpoints are somewhat lower, there can be stopping in the interior.
 
\begin{figure}[t] \centering\includegraphics[scale=0.9]{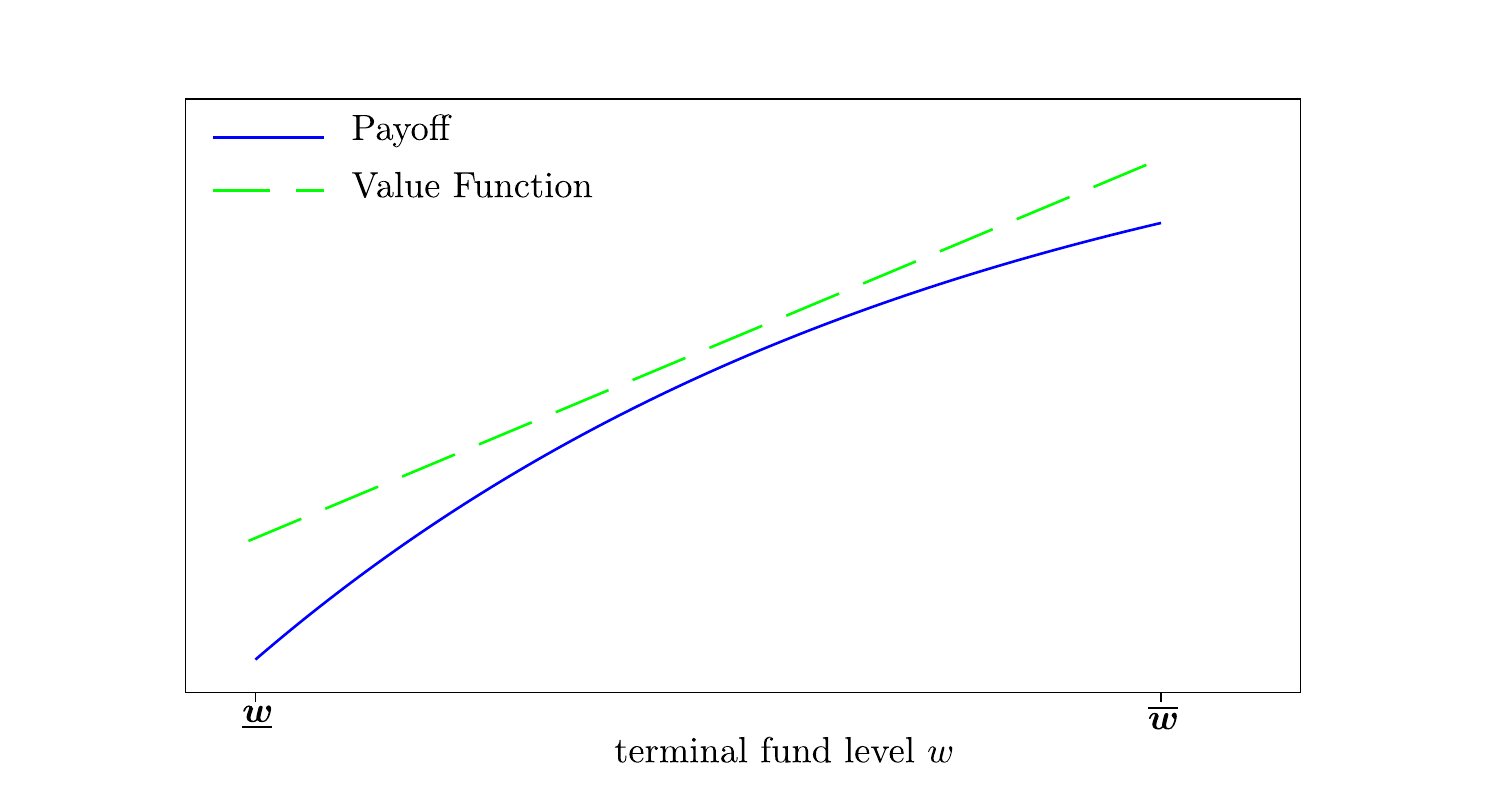}
Figure B
\end{figure}

Figure B illustrates the typical situation for values $(\wl, \wh)$ close enough to $(w_0, w_0)$ that it is beneficial to keep going; the set of such values we call the \textit{continuation region}. Figure C illustrates the situation once $(\wl, \wh)$ has moved sufficiently far from $(w_0, w_0)$. 
\begin{figure}[t] \centering \includegraphics[scale=0.9]{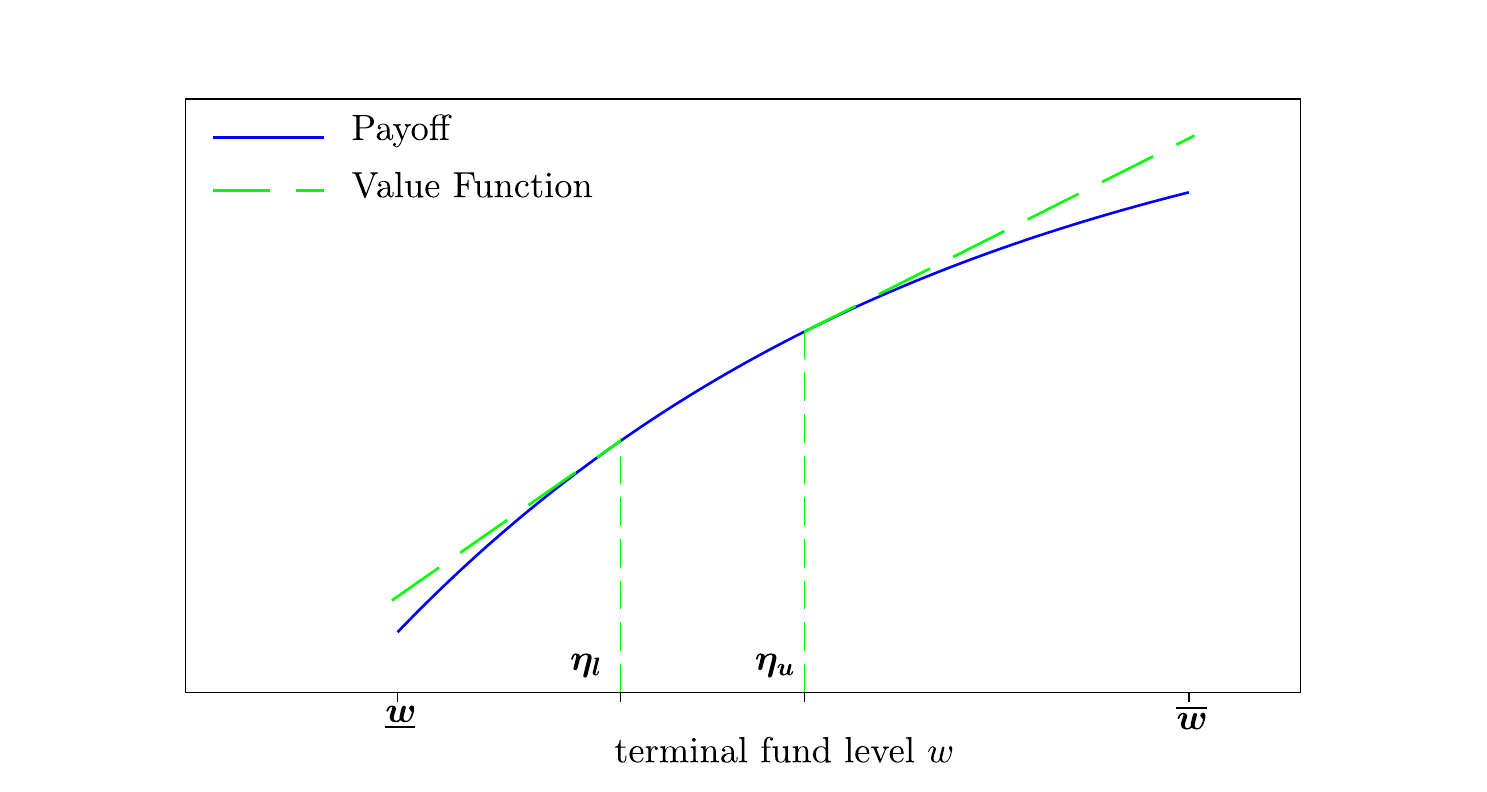} 
Figure C
\end{figure}
Figures \ref{Fig3} and \ref{Fig4} plot out the continuation region and the barriers $\eta_l, \eta_u$ for our first example. Various comments are in order
\begin{itemize}
	\item[(i)] If for some $(\wl, \wh)$ there is optimal stopping at some interior value, then both $\eta_l<\wh+h$ and $\eta_u>\wl-h$. It is not possible to have $\eta_l<\wh+h$ and $\eta_u=\wl-h$, or $\eta_u>\wl-h$ but $\eta_l=\wh+h$. Note that if $\eta_l<\wh+h$ (and therefore $\eta_u>\wl-h$), we always have $\eta_l \le \eta_u$.
	\item[(ii)] In the plots computed, the continuation region is a connected set. In general for an optimal stopping problem with stopping reward $g(\wl, w, \wh)$ this does not need to happen.
	\item[(iii)] In the plots computed, we have the property that if $(\wl, \wh)$ is not in the continuation region, then neither is $(x, y)$ for any $x\le \wl,\ y\ge \wh$. This means that once $(\wl, \wh)$ leaves the continuation region, no further crossing of $[\wl, \wh]$ will happen. So if we first leave the continuation region by an increase of $\wh$, then $\wl$ will not go any lower; we would always choose to stop before that happened. This would be the case of a \textit{successful} fund which has risen in value; the manager will stop only if the fund level falls for enough from the maximum to endanger the gains and we find that actually only a small fall will trigger stopping. If $(\wl, \wh)$ leaves the continuation region by $\wl$ falling, we are seeing an \textit{unsuccessful} fund which has made significant losses. There we see that the stopping barrier is actually quite high; the manager will keep on gambling in the hope of recovering some of the losses and will either gamble to extinction or until enough of the losses have been recovered that he will choose to stop.
\end{itemize}

\section{Conclusions.}\label{S4}
We have taken the problem of a fund manager whose objective is to maximize
the expected utility of his wealth, which is made up of a performance
fee and a management fee. Under certain simplifying assumptions, we argue
that his objective is a function only of the terminal level of the fund,
 and the maximum and minimum levels achieved by the fund. A general argument
 equates the investment problem to a corresponding optimal stopping
 problem for Brownian motion, which we approximate by discretizing the Brownian
 motion to a random walk; in this form, the problem can be solved efficiently
 numerically, and we illustrate the optimal stopping rule with some numerical 
 examples. 
  
 While stopping problems for Brownian motion based on the value and 
 the running maximum are much studied (see Az\'ema \& Yor \cite{ay79} for 
 a seminal contribution), there has been less attention to stopping
 problems involving the value, the running maximum {\em and the running
 minimum} (though see the recent paper of Cox and Ob\l oj \cite{co11} for 
 an important contribution.) The existing literature deals with such 
 questions in the context of finding joint laws for the two (or three)
 variables $\Bh_\tau$, $B_\tau$ (and $\Bl_\tau$) which are extremal in 
 some sense\footnote{See
 Rogers \cite{r93}  where the stochastically largest maximum of a martingale
 whose terminal distribution is specified is shown to be achieved
 by the Az\'ema-Yor construction.}, and the analysis is
  typically quite detailed.  The flavour of the present study is somewhat
  different however, and we readily turn to numerical methods because the
  problem is too complicated to be amenable to analysis.

\section{Appendix} 

 \noindent{\sc Proof of Lemma \ref{L1}.} 
 
First we prove that
\begin{equation*}
	\sup_\theta\, E  F(\wl_1, w_1, \wh_1)  \geq \sup_{T \in {\cal T} }\, E F(\Bl_T, B_T,\Bh_T).
\end{equation*}
 If $T \in {\cal T}$, the process
\begin{equation}
	w_t \equiv B\biggl( \frac{t}{1-t}\; \wedge \; T\biggr)
\end{equation}
may be represented as 
\begin{equation}
	w_t = \int_0^t I_{ \{ s \leqslant  T' \} } \frac{dW_s}{1-s},\label{optstratfromst}
\end{equation}
where $T' = T/(1+T)$ and $W$ is the standard Brownian motion
defined by
\begin{equation}
\int_0^t \frac{dW_s}{1-s} = B\biggl( \frac{t}{1-t}\;\biggr).
\end{equation}
Accordingly, $(\wl_1, w_1, \wh_1) = (\Bl_T, B_T, \Bh_T)$, proving the first inequality.

For the converse inequality, we first notice that
\begin{equation}
\sup_\theta\, E  F(\wl_1, w_1, \wh_1)  =
\sup_{\theta \in \sS}
\, E  F(\wl_1, w_1, \wh_1)=
\sup_\varepsilon \sup_{\theta \in \sS_\varepsilon}
\, E  F(\wl_1, w_1, \wh_1),
\end{equation}
where $\sS$ is the vector space of simple processes
\begin{equation}
\theta = \sum_{j=0}^n Z_j I_{  \{ t_j < t \leqslant  t_{j+1}\}}
\label{simple}
\end{equation}
for some  $ 0 = t_0 < t_1 < \ldots < t_{n+1} = 1$, and 
$Z_j \in L^\infty(\sF_{t_j})$ for all $j$,
and $\sS_\varepsilon = \{ \theta \in \sS : |\theta|
 \geq\varepsilon\}$.  So it will be sufficient to show that 
 whenever we have some $\theta \in \sS_\varepsilon$ then 
 there is some Brownian motion $B$ with canonical
 filtration $(\sB_t)$ and a $(\sB_t)$-stopping time
 $T$  such that 
\begin{equation}
E  F(\wl_1, w_1, \wh_1) = E F(\Bl_T, B_T,\Bh_T).
\label{todo}
\end{equation}
Given $\theta \in \sS_\varepsilon$ of the form \eqref{simple}, 
we form the quadratic variation process 
\begin{equation}
A_t = \int_0^t \theta_s^2 \; ds, 
\end{equation}
and define $B_t = w(\tau_t)$, where $\tau$ is the continuous
strictly-increasing inverse to $A$.  Next define
\begin{equation}
T_k = A_{t_k} = \sum_{j=0}^{k-1} Z_j^2 (t_{j+1} - t_j).
\end{equation}
We claim that for each $k$,
$\sB_s = \sF_{\tau_s}$ for all $0 \leqslant  s \leqslant  T_k$.
 It is clear that 
$\sB_{T_k} \subseteq \sF_{t_k}$, but we shall prove by
induction that equality holds for all $k$. Evidently equality
holds for $k=0$, since both $\sigma$-fields are trivial.
Suppose true up to some value of $k$. Then $\sB(T_k) =
\sF(t_k)$, and so $Z_k$ is $\sB(T_k)$-measurable.
Now for $0\leqslant  s \leqslant  T_{k+1}-T_k$ we have
\begin{equation}
B_{T_k+s}  - B_{T_k} = Z_k\{\,  w(t_k + sZ_k^{-2}) - w(t_k)\,\}
\end{equation}
and therefore we can deduce the path $(w(t_k+u))_{0 \leqslant  u
\leqslant t_{k+1}-t_k}$ from the path of $(B_u)_{0 \leqslant u \leqslant 
T_{k+1}})$, since $Z_k$ is $\sB(T_k)$-measurable.
  This extends the conclusion out to 
$k+1$, and hence for all $k$, and the equality \eqref{todo}
follows from taking $k=n+1$.\\

 \noindent{\sc Proof of Lemma \ref{L2}.} 
 
 For $\h > 0$ we are going to embed a scaled random walk in our 
Brownian motion. Let therefore $\sigma^\h_0=0$ and 
\begin{align*}
	\sigma^\h_{n+1}	= \inf\{t>\sigma^\h_n; \ |B_t - B_{\sigma^\h_n}| = \h\},\quad n \ge 0.
\end{align*}
Then $w^\h_n=B_{\sigma^\h_n},\ n\ge 0$ clearly defines a 
random walk on $B_0 + \h \mathbb Z$ satisfying
\begin{align}
\label{approxbmrw}
		\|(\Bl_t,B_t, \Bh_t) - (\wl^\h_n,w^\h_n,\wh^\h_n)\| \le \h\sqrt{3},
\end{align}
for $\sigma^\h_{n-1} \le t \le \sigma^\h_{n+1},\ n \ge 1$. Any 
$w^\h$-stopping time $\tau$ naturally induces an 
$B$-stopping time $\hat \tau=\sigma^\h_\tau$ with 
$w^\h_\tau=B_{\hat \tau}$, giving us
\begin{align}\label{firstineq}
	V \ge V^\h.
\end{align}
Moreover, for any $B$-stopping time $\tau$,
\begin{align*}
	\tau^\h=\inf\{n \ge 0;\ \sigma^\h_n \ge \tau \}
\end{align*}
defines a random time with 
\begin{align*}
	| w^\h_{\tau^\h}- B_{\tau} | \le \h,\qquad
	| \wh^\h_{\tau^\h}- \Bh_{\tau} | \le \h,\qquad
	| \wl^\h_{\tau^\h}- \Bl_{\tau} | \le \h.
\end{align*}
However, $\tau^\h$ is not necessarily a $w^h$-stopping time. Define $\mathcal G^h_n=\sigma(w^h_j: j\le n)$ and $\mathcal G^h = \sigma(w^h_j: j \ge 0)$. Now let $\pi_n =  P[ \tau \in (\sigma^h_{n}, \sigma^h_{n-1}] | \mathcal G^h]$. Note that
\begin{align*}
	\pi_n =  P[ \tau \in (\sigma^h_{n-1}, \sigma^h_{n}] | \mathcal G^h_n]
\end{align*}
since $\mathcal G^h_n = \sigma( \xi_j: j=1...n)$, with $\xi_j$ IID. Now we give ourselves a uniform random variable $U$, independent of $B$ and add this to every $\mathcal G^h_n$ to form $\tilde{\mathcal G}^h_n = \mathcal G^h_n \vee  \sigma(U)$. If we stop the random walk $w^h_n$ at the $\tilde{\mathcal G}^h$-stopping time
\begin{align}
	S= \inf\{n: \sum_{i=1}^n \pi_i > U \}.
\end{align}
we find that $w^h_S$ has the same law as $w^h_{\tau^h}$ and then
\begin{align}\label{secondineq}
	V^\h \ge V - \psi(h\sqrt 3).
\end{align}
  Combining (\ref{firstineq}) and (\ref{secondineq}) proves the desired result.

\bibliography{InvStop}

\begin{thebibliography}{1}

\bibitem{ay79}
J.~Az\'ema and M.~Yor.
\newblock Une solution simple au proml\`eme de skorokhod.
\newblock {\em S\'eminaire de probabilit\'es}, 13:90--115, 1979.

\bibitem{co11}
A.~M.~G. Cox and J.~Ob\l oj.
\newblock Robust hedging of double touch barrier options.
\newblock {\em SIAM Journal on Financial Mathematics}, 2:141--182, 2011.

\bibitem{dk96}
P.~H. Dybvig and H.-K. Koo.
\newblock Investment with taxes.
\newblock {\em Working paper}, 1996.

\bibitem{gir03}
W.~N. Goetzmann, Jr. J.~E.~Ingersoll, and S.~A. Ross.
\newblock High-water marks and hedge fund management contracts.
\newblock {\em The Journal of Finance}, 4:1685--1717, 2003.

\bibitem{go10}
P.~Guasoni and J.~Ob\l oj.
\newblock The incentives of performance fees and high water marks.
\newblock {\em Preprint}, 2010.

\bibitem{r93}
L.~C.~G. Rogers.
\newblock The joint law of the maximum and terminal value of a martingale.
\newblock {\em Probability Theory and Related Fields}, 95:451--466, 1993.

\end{thebibliography}
\bibliographystyle{plain}

\begin{figure}[h!] 
	\centering 
	\includegraphics[scale=0.75]{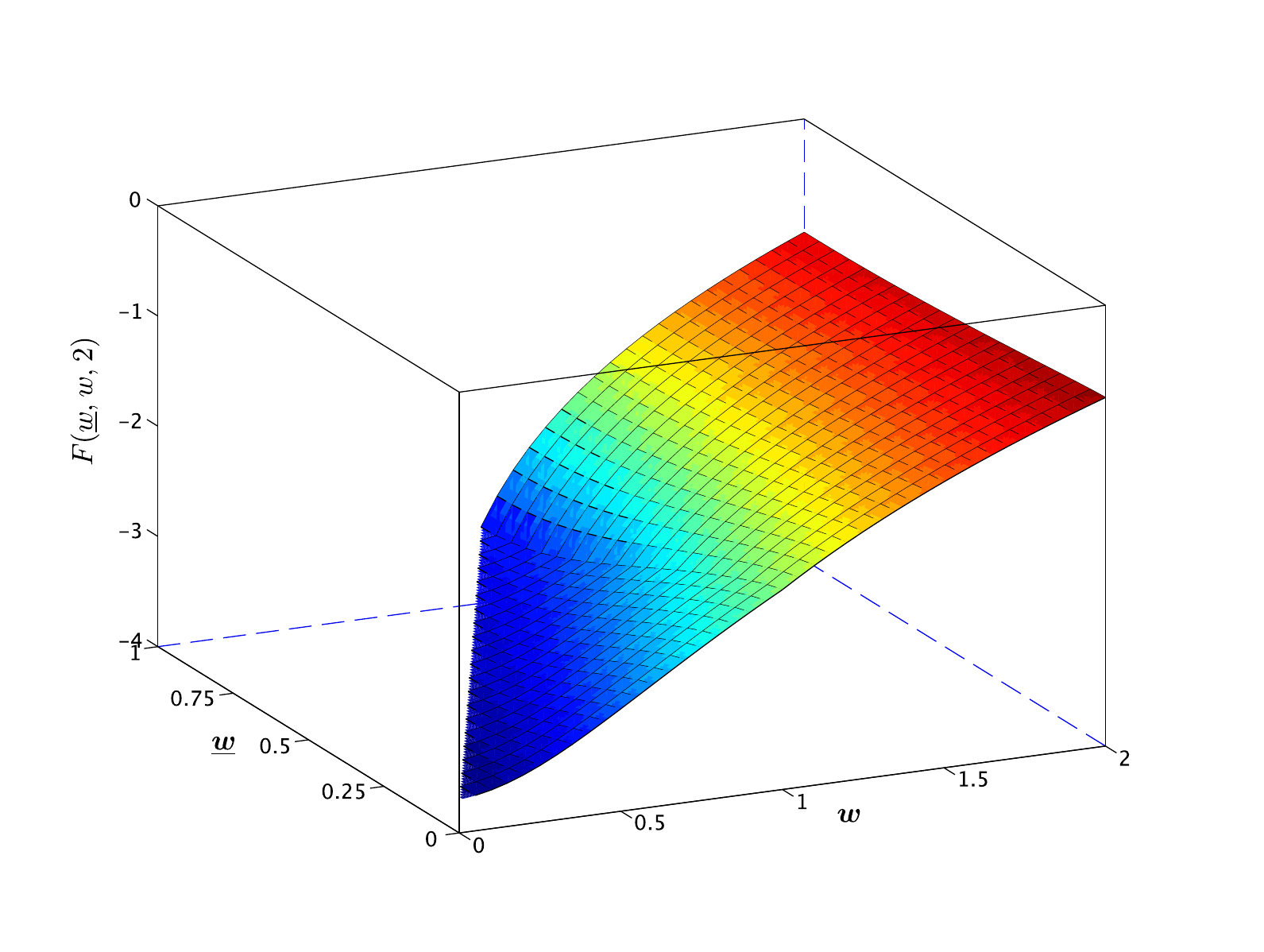}
	\caption{$F(\wl, w, \wh)$ for fixed $\wh\!=\!2$: S-shaped in $w$ and decreasing in $\wl$.}
	\label{Fig1}

  	\includegraphics[scale=0.75]{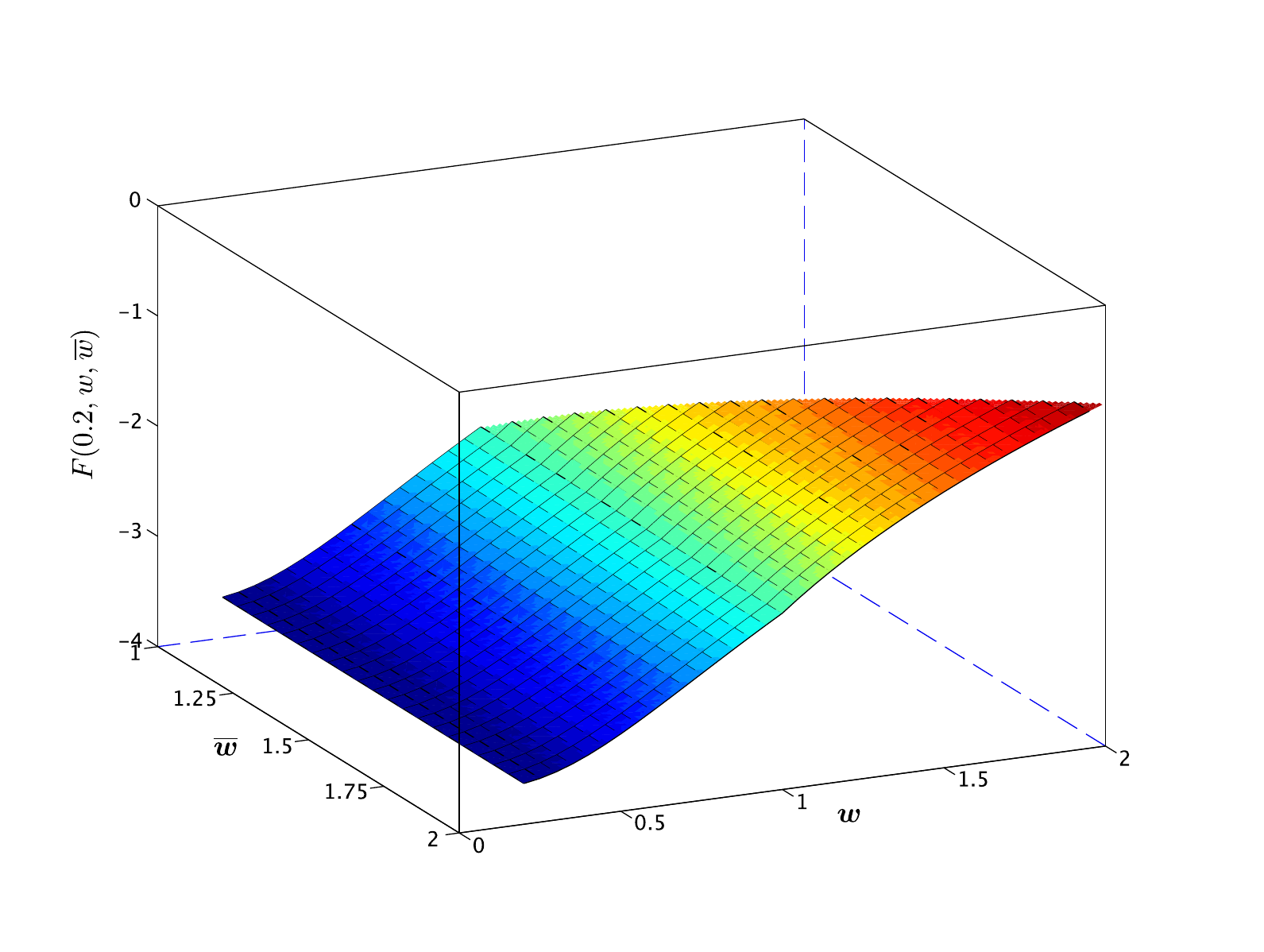}
	\caption{$F(\wl, w, \wh)$ for fixed $\wl\!=\!0.2$: S-shaped in $w$ and slightly increasing in $\wh$.}
	\label{Fig2}
\end{figure}

\begin{figure}[h!]
  	\centering
  	\includegraphics[scale=0.75]{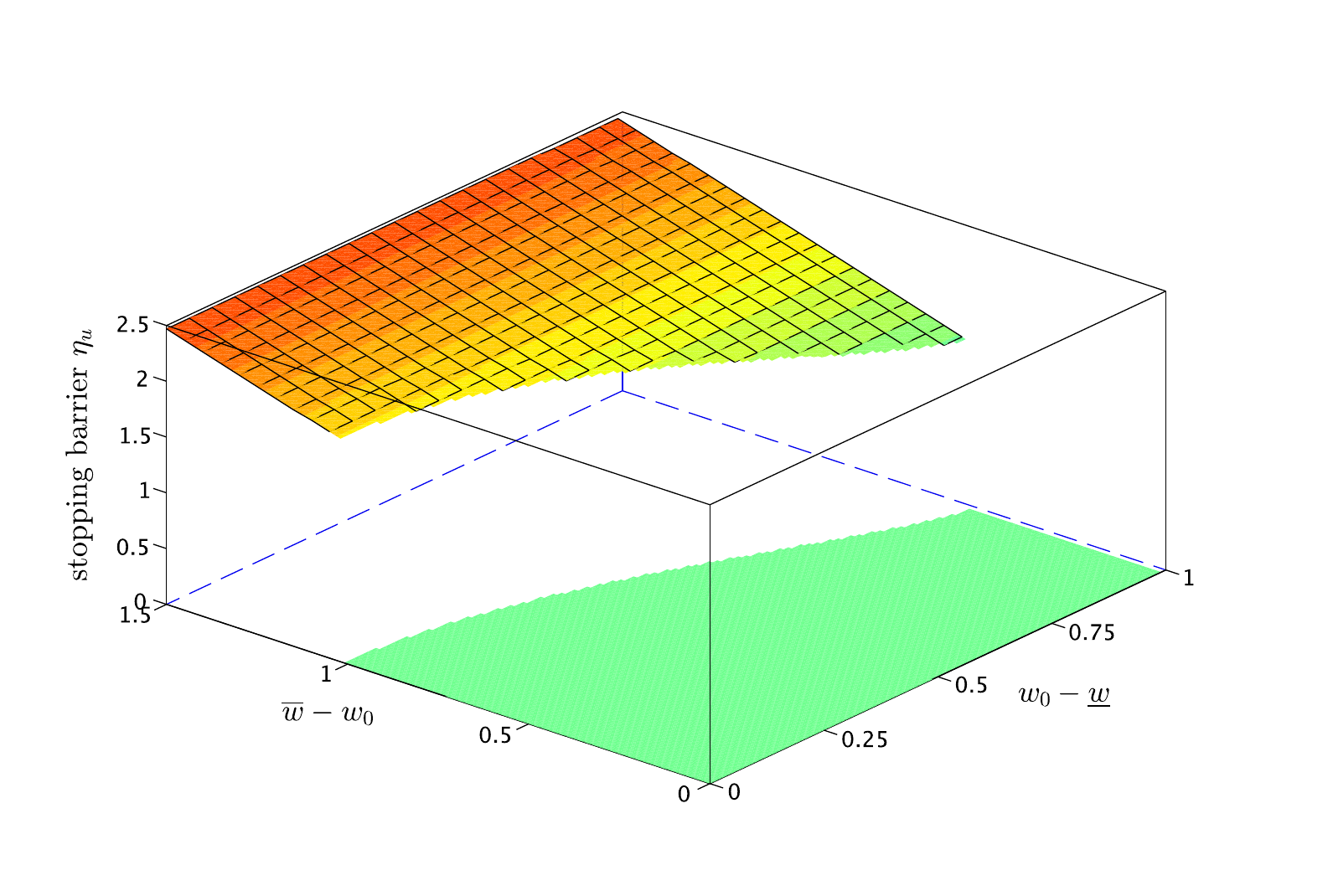}
	\caption{Surface plot of the upper stopping barrier $\eta_l$ that becomes relevant when we leave the continuation region (grey, flat) by decreasing $\wl$. }
	\label{Fig3}

  	\includegraphics[scale=0.75]{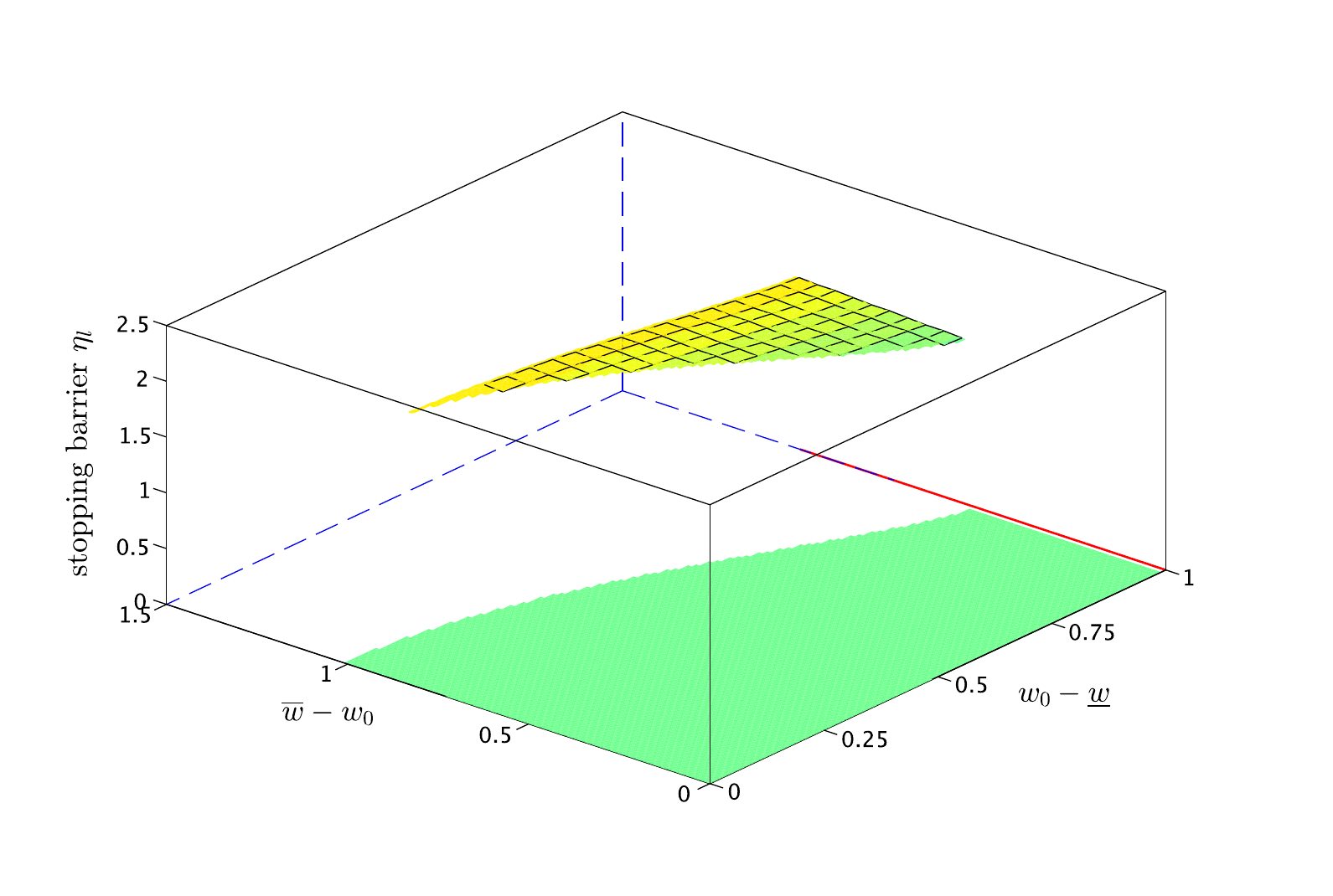}
	\caption{Surface plot of the lower stopping barrier $\eta_u$ that becomes relevant when we leave the continuation region (grey, flat) by increasing $\wh$. }
	\label{Fig4}
\end{figure}

\end{document}